\documentclass[11pt,epsf]{article}
\usepackage{graphicx}
\usepackage{color}
\usepackage{amsmath}
\usepackage{amsfonts}
\usepackage{mathrsfs}
\usepackage{mathtools}
\usepackage{amssymb}
\usepackage{float}
\usepackage{url}
\usepackage{verbatim}
\usepackage{dsfont}
\usepackage{enumitem}
\usepackage{layout}
\usepackage{comment}

\newcommand {\hbx} {\mbox{\boldmath $\hat{x}$}}

\newcommand {\hx} {\hat{x}}
\newcommand {\hX} {\hat{X}}

\newcommand{\tx}{{\tilde{x}}}
\newcommand {\dfn} {\stackrel{\Delta} {=}}

\newcommand {\reals} {{\rm I\!R}}

\newcommand {\bE} {\mbox{\boldmath $E$}}

\newcommand{\calA}{{\cal A}}
\newcommand{\calB}{{\cal B}}

\newcommand{\calE}{{\cal E}}

\newcommand{\calI}{{\cal I}}

\newcommand{\calS}{{\cal S}}

\newcommand{\calX}{{\cal X}}

\newcommand{\calZ}{{\cal Z}}
\allowdisplaybreaks

\begin{document}
\thispagestyle{empty}
\title{Lossy Compression of Individual Sequences Revisited:\\
Fundamental Limits of Finite-State Encoders}
\author{Neri Merhav
}
\date{}
\maketitle

\begin{center}
The Andrew \& Erna Viterbi Faculty of Electrical Engineering\\
Technion - Israel Institute of Technology \\
Technion City, Haifa 32000, ISRAEL \\
E--mail: {\tt merhav@ee.technion.ac.il}\\
\end{center}
\vspace{1.5\baselineskip}
\setlength{\baselineskip}{1.5\baselineskip}

\begin{abstract}
We extend Ziv and Lempel's model of finite-state encoders to the
realm of lossy compression of individual sequences. In particular, the model
of the encoder includes a finite-state reconstruction codebook followed by
an information lossless finite-state encoder that compresses the
reconstruction codeword with no additional distortion. We first derive two
different lower
bounds to the compression ratio that depend on the number of states of the
lossless encoder.
Both bounds are asymptotically achievable by
conceptually simple coding schemes. We then show that when the number of
states of the lossless encoder is large enough in terms of the reconstruction
block-length, the performance can be improved, sometimes significantly so.
In particular, the improved performance is achievable using a random-coding
ensemble that is universal, not only in terms of the source sequence, but also in terms
of the distortion measure.
\end{abstract}

\section{Introduction}

\begin{figure}[h]
\hspace*{.3cm}\includegraphics[width=.95\textwidth]{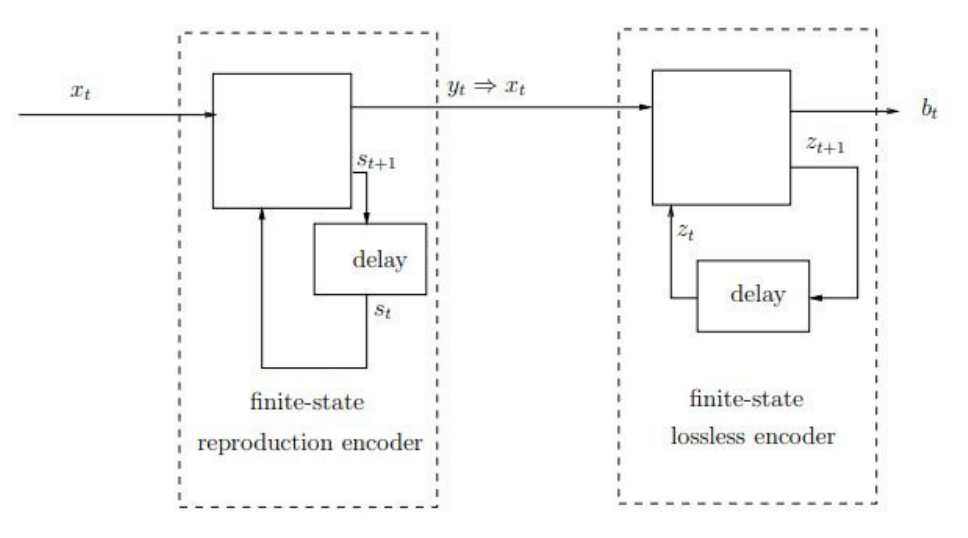}
\caption{\small Finite-state reproduction encoder followed by a
finite-state
lossless encoder.}
\label{fslossyg}
\end{figure}

We revisit the classical domain of rate-distortion coding applied to
finite-alphabet sequences, focusing on a prescribed distortion function
\cite{Berger71}, \cite[Chap.\ 10]{CT06}, \cite[Chap.\ 9]{Gallager68},
\cite{Gray90}, \cite[Chaps.\ 7,8]{VO79}. Specifically, our attention is
directed towards encoders comprising finite-state reproduction encoders
followed by information-lossless finite-state encoders that compress
reproduction sequences without introducing additional distortion (see Fig.\
\ref{fslossyg}). In essence, our principal findings are in establishing two
asymptotically achievable lower bounds for the optimal compression ratio of an
individual source sequence of length $n$, utilizing any finite-state encoder
with the aforementioned structure, where the lossless encoder possesses $q$
states. These lower
bounds can both be conceptualized as the individual-sequence counterparts to the
rate-distortion function of the given source sequence, akin to the lossless
finite-state compressibility of a source sequence serving as the
individual-sequence analogue of entropy. However, before delving into the
intricacies of our results, a brief overview of the background is warranted.

Over the past several decades, numerous research endeavors have been spurred
by the realization that source statistics are seldom, if ever, known in
practical scenarios. Consequently, these efforts have been dedicated to the
pursuit of universal coding strategies that remain independent
of unknown statistics while asymptotically approaching lower bounds, such as
entropy in lossless compression or the rate-distortion function in the case of
lossy compression, as the block length extends indefinitely. Here, we offer a
succinct and non-exhaustive overview of some pertinent earlier works.

In the realm of lossless compression, the field of universal source coding has
achieved a high level of sophistication and maturity. Davisson's seminal work
\cite{Davisson73} on universal-coding redundancies has introduced the pivotal
concepts of weak universality and strong universality, characterized by
vanishing maximin and minimax redundancies, respectively. This work has also
elucidated the link between these notions and the capacity of the 'channel'
defined by the family of conditional distributions of the data to be
compressed, given the index or parameter of the source in the class
\cite{Gallager76}, \cite{Ryabko79}, \cite{DLG80}.
For numerous parametric source classes encountered in practice, the minimum
achievable redundancy of universal codes is well-established to be dominated
by $\frac{k\log n}{2n}$, where $k$ denotes the number of degrees of freedom of
the parameter, and $n$ is the block length \cite{KT81}, \cite{Shtarkov87},
\cite{BRY98}, \cite{YB99}. Davisson's theory gives rise to a central idea of
constructing a Shannon code based on the probability distribution of the data
vector with respect to a mixture, incorporating a certain prior function, of
all sources within the class.
Rissanen, credited with the invention of the minimum description length (MDL)
principle \cite{Rissanen78}, established a converse to a coding theorem in
\cite{Rissanen84}. This theorem asserts that asymptotically, no universal code
can achieve redundancy below $(1-\epsilon)\frac{k\log n}{2n}$ with a possible
exception of sources from a subset of the parameter space, the volume of
which diminishes as $n\to\infty$ for every positive $\epsilon$. Merhav and
Feder \cite{MF95} generalized this result to more extensive classes of
sources, substituting the term $\frac{k\log n}{2n}$ with the capacity of the
aforementioned 'channel'. Subsequent studies have further refined redundancy
analyses and contributed to ongoing developments in the field.

In the broader domain of universal lossy compression, the theoretical
landscape is regrettably not as sharply defined and well-developed as in the
lossless counterpart. In this study, we narrow our focus to a specific class
known as $d$-semifaithful codes \cite{OS90} codes that fulfill the distortion
requirement with probability one.
Zhang, Yang, and Wei \cite{ZYW97} have demonstrated a notable contrast with
lossless compression, establishing that, even when source statistics are
perfectly known, achieving redundancy below $\frac{\log n}{2n}$ in the lossy
case is impossible, although $\frac{\log n}{n}$ is attainable. The absence of
source knowledge imposes a cost in terms of enlarging the multiplicative
constant associated with $\frac{\log n}{n}$. Yu and Speed \cite{YS93}
established weak universality, introducing a constant that grows with the
cardinalities of the source and reconstruction alphabets \cite{SP21}.
Ornstein and Shields \cite{OS90} delved into universal $d$-semifaithful coding
for stationary and ergodic sources concerning the Hamming distortion measure,
demonstrating convergence to the rate-distortion function with probability
one. Kontoyiannis \cite{Kontoyiannis00} made several noteworthy contributions.
Firstly, a central limit theorem (CLT) with a $O(1/\sqrt{n})$ redundancy term,
featuring a limiting Gaussian random variable with constant variance.
Secondly, the law of iterated logarithm (LIL) with redundancy proportional to
$\sqrt{\log(\log n)/n}$ infinitely often with probability one. A
counter-intuitive conclusion from \cite{Kontoyiannis00} is the priceless
nature of universality under these CLT and LIL criteria.
In \cite{KZ02}, optimal compression is characterized by the negative logarithm
of the probability of a sphere of radius $nD$ around the source vector
with respect to the distortion measure, where $D$ denotes the allowed per-letter
distortion. The article also introduces the concept of random coding ensemble
with a probability distribution given by a mixture of all distributions in a
specific class.
In two recent articles, Mahmood and Wagner \cite{MW22a}, \cite{MW22b} have
delved into the study of $d$-semifaithful codes that are strongly universal
concerning both the source and the distortion function.
The redundancy rates in \cite{MW22a} behave like $\frac{\log n}{n}$
but with different
multiplicative constants. Other illuminating results regarding a special distortion measure
are found in \cite{Sholomov06}.

A parallel path of research in the field of universal lossless and lossy
compression, spearheaded by Ziv, revolves around the individual-sequence
approach. In this paradigm, no assumptions are made about the statistical
properties of the source. The source sequence to be compressed is
treated as an arbitrary deterministic (individual) sequence, but instead
limitations are imposed on the implementability of the encoder and/or decoder using
finite-state machines. This approach notably encompasses the widely celebrated
Lempel-Ziv (LZ) algorithm \cite{Ziv78}, \cite{ZL78}, \cite{Potapov04}, along
with subsequent advancements broadening its scope to both lossy compression
with and without side information \cite{MZ06}, \cite{Ziv84}, as well as joint
source-channel coding \cite{me22}, \cite{Ziv80}.
In the lossless context, the work in \cite{WMF94} establishes an
individual-sequence analogue akin to Rissanen's result, where the expression
$\frac{k\log n}{2n}$ continues to denote the best achievable redundancy.
However, the primary term in the compression ratio is the empirical entropy of
the source vector, deviating from the conventional entropy in the
probabilistic setting. The converse bound presented in \cite{WMF94} is
applicable to the vast majority of source sequences within each type, echoing
the analogy with Rissanen's framework concerning the majority of the parameter
space. It is noteworthy that this converse result retains a semblance of the
probabilistic setting, as asserting the relatively small number of exceptional
typical sequences is equivalent to assuming a uniform distribution across the
type and asserting a low probability of violating the bound.
Conversely, the achievability result in \cite{WMF94} holds pointwise for every
sequence. A similar observation applies to \cite{me-univdis}, where
asymptotically pointwise lossy compression was established concerning
first-order statistics (i.e., ``memoryless'' statistics), emphasizing
distortion-universality, akin to the focus in \cite{MW22a} and \cite{MW22b}.
A similar fusion of the individual-sequence setting and the probabilistic
framework is evident in \cite{me23} concerning universal rate-distortion
coding. However, akin to the approach in \cite{me-univdis}, there is no
constraint on finite-state encoders/decoders as in \cite{WMF94}. Notably, the
converse theorem in \cite{me23} states that for any variable-rate code and any
distortion function within a broad class, the vast {\em majority} of reproduction
vectors representing source sequences of a given type (of any fixed order)
must exhibit a code length essentially no smaller than the negative logarithm
of the probability of a ball with a normalized radius $D$ (where $D$ denotes
the allowed per-letter distortion). This ball is centered at the specified
source sequence, and the probability is computed with respect to a universal
distribution proportional to $2^{-LZ(\hbx)}$, where $LZ(\hbx)$ denotes the
code length of the LZ encoding of the reproduction vector $\hbx$.

The emphasis on the term ``majority'' in the preceding paragraph, as highlighted
earlier, necessitates clarification. It should be noted that in the absence of
constraints on encoding memory resources, such as the finite-state machine
model mentioned earlier, there cannot exist any meaningful lower bound that
universally applies to each and every individual sequence. The rationale is
straightforward: for any specific individual source sequence, it is always
possible to devise an encoder compressing that sequence to a single bit (even
losslessly). For instance, by designating the bit '0' as the compressed
representation of the given sequence and appending the bit '1' as a header to
the uncompressed binary representation of any other source sequence. In this
scenario, the compression ratio for the given individual sequence would be
$1/n$, dwindling to zero as $n$ grows indefinitely. It is therefore clear
that any non-trivial lower bound that universally applies to
{\em every} individual source sequence at the same time, necessitates reference to a class of encoders/decoders
equipped with constrained resources, such as those featuring a finite number
of states.

In this work, we consider lossy compression of individual
source sequences using finite-state
encoders whose structure is as follows: Owing to the fact that, without loss of
optimality, every lossy encoder can be presented as a cascade of a
reproduction encoder and a lossless (or ``noiseless'') encoder (see, e.g.,
\cite[discussion around Fig.\ 1]{NG82}), we consider a class of lossy encoders
that can be implemented as a cascade of a finite-state reproduction encoder
and a finite-state lossless encoder, see Fig.\ \ref{fslossyg}. The finite-state reproduction encoder
model is a generalization of the well-known finite-state vector quantizer
(FSVQ), see, e.g., \cite{FGOD85}, \cite[Chap.\ 14]{GG92}. It is committed to
produce reproduction vectors of dimension $k$ in response to source vectors of
dimension $k$, while complying with the distortion constraint for every such
vector. The finite-state lossless encoder is the same as in \cite{ZL78}.
The number of states of the reproduction encoder can be assumed very large
(in particular, even large enough to store many recent input blocks).
Both the dimension, $k$, and the number of states, $q$, of the lossless encoder 
are assumed small compared to the total length, $n$, of the source sequence to
be compressed, similarly as in \cite{ZL78} (and other related works), where
the regime $q\ll n$ is assumed too. 

One of our main messages in this
work is that it is important also how $q$
and $k$ related to each other, and not only how to they both relate to $n$. 
If $q$ is large in terms of $k$, one can do
much better than if it is small. Accordingly, we first derive two different
lower bounds to the compression ratio under the assumption that $q\ll k$,
which are both asymptotically achievable by conceptually simple schemes that,
within each $k$-block, seek the most compressible $k$-vector within a ball of
`radius' $kD$ around the source block. We then compare it to the ensemble performance
of a universal coding scheme that can be implemented when $q$ is exponential in $k$.
The improvement can sometimes be considerably large. The universality of the
coding scheme is two-fold: both in the source sequence to be compressed and in
the distortion measure in the sense that the order of codewords within the
typical codebook (which affects the encoding of their indices) is asymptotically
optimal no matter which distortion measure is used (see \cite{me23} for
discussion of this property).
The intuition behind this improvement is that when $q$ is exponential in $k$,
the memory of the lossless encoder is large enough to store entire input
blocks and thereby exploit
the sparseness of the reproduction codebook in the
space of $k$-dimensional vectors with components in the reproduction alphabet.
The asymptotic achievability of the lower bound will rely on the direct coding
theorem of \cite{me23}.

Bounds on both lossless and lossy compression of individual sequences using
finite-state encoders and decoders have been explored in previous works,
necessitating a contextualization of the present work. As previously
mentioned, the cases of (almost) lossless compression were examined in
\cite{Ziv78}, \cite{ZL78}, and \cite{Ziv84}. In \cite{Ziv80}, the lossy case
was considered, incorporating both a finite-state encoder and a finite-state
decoder in the defined model. However, in the proof of the converse part, the
assumption of a finite-state encoder was not essential; only the finite number
of states of the decoder was required. The same observation holds for
\cite{me14}. In a subsequent work, \cite{me22}, the finite number of states
for both the encoder and decoder was indeed utilized. This holds true for
\cite{MZ06} as well, where the individual-sequence analogue of the Wyner-Ziv
problem was investigated with more restrictive assumptions on the structure of
the finite-state encoder.
In contrast, the current work restricts only the encoder to be a finite-state
machine, presenting a natural generalization of \cite{ZL78} to the lossy case.
Specifically, one of our achievable lower bounds can be regarded as an extension of
the compressibility bound found in \cite[Corollary 1]{ZL78} to the lossy
scenario. It is crucial to note that, particularly in the lossy case, it is
more imperative to impose limitations on the encoder than the decoder, as
encoding complexity serves as the practical bottleneck. Conversely, for
deriving converse bounds, it is stronger and more general not to impose any
constraints on the decoder.

The outline of this paper is as follows. In Section 2, we establish notation,
as well definitions, and spell out the objectives.
In Section 3, we derive the main results and discuss them.
Finally, in Section 4 we summarize the main contributions of this work and
make some concluding remarks.

\section{Notation, Definitions and Objectives}
\label{ndo}

Throughout the paper, random variables will be denoted by capital
letters, specific values they may take will be denoted by the
corresponding lower case letters, and their alphabets
will be denoted by calligraphic letters. Random
vectors and their realizations will be denoted,
respectively, by capital letters and the corresponding lower case letters,
both in the bold face font. Their alphabets will be superscripted by their
dimensions. The source vector of length $n$, $(x_1,x_2,\ldots,x_n)$, with
components, $x_i$, $i=1,2,\ldots,n$, from a
finite-alphabet, $\calX$, will be denoted by $x^n$. The set of all such
$n$-vectors will be denoted by
$\calX^n$, which is the $n$--th order Cartesian power of $\calX$. 
Likewise, a reproduction vector of length $n$, $(\hx_1,\ldots,\hx_n)$, with 
components, $\hx_i$, $i=1,\ldots,n$, from a
finite-alphabet, $\hat{\calX}$, will be denoted by $\hx^n\in\hat{\calX}^n$. 
The notation $\hat{\calX}^*$ will be used to designate the set of all
finite-length strings of symbols from $\hat{\calX}$. 

For $i\le j$, the notation $x_i^j$ will be used to denote the substring
$(x_i,x_{i+1},\ldots,x_j)$. For $i=1$, the subscript `1' will be omitted, and
so, the shorthand notation of $(x_1,x_2,\ldots,x_n)$ would be $x^n$. Similar
conventions will apply to other sequences.
Probability distributions will be denoted by the letter $P$ or $Q$ with
possible subscripts, depending on the context.
The probability of an event $\calA$ will be denoted by $\mbox{Pr}\{\calA\}$,
and the expectation
operator with respect to (w.r.t.) a probability distribution $P$ will be
denoted by
$\bE\{\cdot\}$.
The logarithmic function, $\log x$, will be understood to be defined to the
base 2. Logarithms to the base $e$ will be denote by $\ln$.
Let $d:\calX\times\hat{\calX}\to\reals$ be a given distortion function between
source symbols and reproduction symbols. The distortion between vectors will be
defined additively as $d(x^n,\hx^n)=\sum_{i=1}^nd(x_i,\hx_i)$ for every
positive integer, $n$, and every $x^n\in\calX^n$, $\hx^n\in\hat{\calX}^n$.

Consider the encoder model depicted in Fig.\ \ref{fslossyg}, 
which is a cascade of a finite-state reproduction encoder (FSRE) and a
finite-state lossless encoder (FSLE). This encoder
is fully determined by the set
$E=(\calX,\hat{\calX},\calS,\calZ,u,v,f,g,k)$, where $\calX$ is the source input
alphabet of size $\alpha$,
$\hat{\calX}$ is the reproduction alphabet of size $\beta$, $\calS$ is a set of
FSRE states,
$\calZ$ is a set of
FSLE states of size $q$,
$u$ and $v$ are functions that define the FSRE, 
$f$ and $g$ are functions that define the FSLE
(both to be defined
shortly), and $k$ is a positive integer that designates
the basic block length within which the distortion constraint must be kept, as
will be described shortly. The number of states, $|\calS|$, of the
FSRE may be assumed arbitrarily large (as the lower bounds to be derived will
actually be independent of this number). In particular, it can be assumed to be large enough
to store several recent input $k$-blocks.

According to this encoder model, the input, $x_t\in\calX$,
$t=1,2,\ldots$,
is fed sequentially into the FSRE, which goes through a sequence of states
$s_t\in\calS$, and produces an output sequence, $y_t\in\hat{\calX}^*$ of
variable-length
strings of symbols from $\hat{\calX}$, with the possible inclusion of
the empty symbol, $\lambda$, of length zero. Referring to Fig.\
\ref{fslossyg}, the FSRE is defined by
the recursive equations,
\begin{eqnarray}
y_t&=&u(x_t,s_t)\\
s_{t+1}&=&v(x_t,s_t),
\end{eqnarray}
for $t=1,2,\ldots$, where the initial state, $s_1$, is assumed to be some
fixed member of $\calS$.\\

\noindent
{\bf Remark 1.} The above defined model of the FSRE has some resemblance to the well known
model of the finite-state vector quantizer (FSVQ) \cite{FGOD85}, \cite[Chap.\
14]{GG92}, but it is in fact, considerably more general than
the FSVQ. Specifically, the FSVQ works as follows. At each time instant $t$, it receives a
source-vector $x_t$ and outputs a finite-alphabet variable, $u_t$,
while updating its internal state $s_t$. The encoding function is
$u_t=a(x_t,s_t)$ and the next-state function is $s_{t+1}=\phi(u_t,s_t)$.
Note that state evolves in response to $\{u_t\}$ (and not $x_t$), so that the decoder would be able
to maintain its own copy of $\{s_t\}$.
At the decoder, the reproduction is generated according to
$\hx_t=b(u_t,s_t)$ and the state is updated again using
$s_{t+1}=\phi(u_t,s_t)$. By cascading the FSVQ encoder and its decoder, one
obtains a system with input $x_t$ and output $\hx_t$, which is basically
a special case of our FSRE with the functions $u$ and $v$ being given by
$u(x,s)=b(a(x,s),s)$ and
$v(x,s)=\phi(a(x,s),s)$. $\Box$\\

As described above, given an input block of length $k$, $(x_1,x_2,\ldots,x_k)$, the FSRE generates
a corresponding output block, $(y_1,y_2,\ldots,y_k)$, while traversing a
sequence of states $(s_1,\ldots,s_n)$. The FSRE must be
designed in such a way that the total length of the concatenation of the
(non-empty) variable-length strings,
$y_1,y_2,\ldots,y_k$, is equal to $k$ as well.
Accordingly, given
$(y_1,y_2,\ldots,y_k)$, let $(\hx_1,\hx_2,\ldots,\hx_k)$ denote the
corresponding vector of reproduction symbols from $\hat{\calX}$, which forms the output of
the FSRE. This formal transformation from $y^k$ to $\hx^k$ is designated by
the expression $y_t\Rightarrow\hx_t$ in Fig.~\ref{fslossyg}.\\

\noindent
{\bf Example 1.}
Let $\calX=\hat{\calX}=\{a,b,c\}$, and suppose that the FSRE is a block code
of length $k=5$. Suppose also that $x^5=(a,a,b,c,c)$ and
$y^5=(\lambda,\lambda,\lambda,\lambda,
\mbox{'aabbc'})$. Then, $\hx^5=(a,a,b,b,c)$. The current state, in this case, is simply
the contents of the input, starting from the beginning of the current block and
ending at the current input symbol. Accordingly, the encoder idles until the
end of the input block, and then it produces the full output block. $\Box$\\

The parameter $k$ of the encoder $E$ is the length of the basic block that
is associated with the
distortion constraint. 
For a given input alphabet $\calX$, reconstruction alphabet $\hat{\calX}$, and
distortion function $d$, 
we denote by $\calE(q,k,D)$ the class of all finite-state encoders of
the above described structure, whose number of FSLE states is $q$,
the dimension of the FSRE is $k$,
and $d(x^k,\hx^k)\le kD$ for every 
$x^k\in\calX^k$. For future use, we also define the `ball'
\begin{equation}
\calB(x^k,D)=\{\hx^k:~d(x^k,\hx^k)\le kD\}.
\end{equation}

\noindent
{\bf Remark 2.} Note that the role of the state variable, $s_t$, might not be only to store
information from the past of the input, but possibly also to maintain the distortion
budget within each $k$-block. At each time instant $t$, the state can be used to
update the remaining distortion allowed until the end of the current
$k$-block. For example, if the entire allowed distortion budget, $kD$, has
already been exhausted before the current $k$-block has ended, then in the remaining
part of the current block, the encoder must carry on losslessly, that is, it
must produce reproduction symbols
that incur zero-distortion relative to the corresponding source symbols.
$\Box$\\

The FSLE is defined similarly as in \cite{ZL78}. 
Specifically, the output of the FSRE, $\hx_t\in\hat{\calX}$,
$t=1,2,\ldots$,
is fed sequentially into the FSLE, which in turn goes through a sequence of states
$z_t\in\calZ$, and produces an output sequence, $b_t\in\{0,1\}^*$ of
variable-length binary
strings, with the possible inclusion of
the empty symbol, $\lambda$, of length zero. Accordingly,
the FSLE implements
the recursive equations,
\begin{eqnarray}
b_t&=&f(\hx_t,z_t)\\
z_{t+1}&=&g(\hx_t,z_t),
\end{eqnarray}
for $t=1,2,\ldots$, where the initial state, $z_1$, is assumed to be some fixed member
of $\calZ$.

With a slight abuse of notation, we adopt the 
extended use of encoder functions $u$, $v$, $f$ and $g$, to designate output sequences and final
states, which result from 
the corresponding initial states and inputs. We use the notations $u(s_1,x^n)$, $v(s_1,x^n)$, $f(z_1,u(s_1,x^n))$ and 
$g(z_1,u(s_1,x^n))$ for $\hx^n$, $s_{n+1}$, $b^n$, and $z_{n+1}$,
respectively. We assume the FSLE to be information lossless,
which is defined, similarly as in \cite{ZL78}, as follows.
For every $(z_1,s_1)\in\calZ\times\calS$, every positive integer
$n$, and every $x^n\in\calX^n$, the triple
$(z_1,f(z_1,u(s_1,x^n)),g(z_1,u(s_1,x^n)))$ uniquely determines
$\hx^n$. 

Given an encoder
$E=(\calX,\hat{\calX},\calS,\calZ,u,v,f,g,k)\in\calE(q,k,D)$,
and a source string $x^n$, where $n$ is divisible by $k$, the compression ratio of $x^n$ by $\calE$ is
defined as
\begin{equation}
\rho(x^n;E)=\frac{L(b^n)}{n},
\end{equation}
where $L(b^n)=\sum_{t=1}^n\ell(b_t)$, $\ell(b_t)$ being the length (in bits)
of the binary string $b_t$.
Next, define
\begin{equation}
\rho(x^n;\calE(q,k,D))=\min_{E\in\calE(q,k,D)}\rho(x^n;E).
\end{equation}

Our main objective is derive bounds to
$\rho(x^n;\calE(q,k,D))$ for large $k$ and $n\gg k$, with special interest in the
case where $q$ is large enough (in terms of $k$), but still fixed
independent of $n$, so that the FSLE could take
advantage of the fact that not necessarily
every $\hx^n\in\hat{\calX}^n$ can be obtained as an output of the given FSRE.
In particular, a good FSLE with long memory should exploit the sparseness of the
reproduction codebook relative to the entire space of $k$-vectors in
$\hat{\calX}^k$. 

\section{Lower Bounds}
\label{lb}

For the purpose of presenting both the lower bounds and the
achievability, we briefly review a few
terms and facts concerning the 1978 version of Lempel-Ziv algorithm (a.k.a.\ the LZ78
algorithm) \cite{ZL78}.
The incremental parsing procedure of the LZ78 algorithm is a procedure of
sequentially parsing a vector, $\hx^k\in\hat{\calX}^k$, such that each new
phrase is the shortest
string that has not been encountered before as a parsed phrase, with the
possible exception of the last phrase, which might be incomplete. For example,
the incremental parsing of the vector $\hx^{15}=\mbox{abbabaabbaaabaa}$ is
$\mbox{a,b,ba,baa,bb,aa,ab,aa}$. Let $c(\hx^k)$ denote the
number of phrases in $\hx^k$ resulting from the incremental parsing procedure
(in the above example, $c(\hx^{15})=8$).
Let $LZ(\hx^k)$ denote the
length of the LZ78 binary compressed code for $\hx^k$.
According to
\cite[Theorem 2]{ZL78},
\begin{eqnarray}
\label{lz-clogc}
LZ(\hx^k)&\le&[c(\hx^k)+1]\log\{2\beta[c(\hx^k)+1]\}\nonumber\\
&=&c(\hx^k)\log[c(\hx^k)+1]+c(\hx^k)\log(2\beta)+\log\{2\beta[c(\hx^k)+1]\}\nonumber\\
&=&c(\hx^k)\log c(\hx^k)+c(\hx^k)\log\left[1+\frac{1}{c(\hx^k)}\right]+
c(\hx^k)\log(2\beta)+\log\{2\beta[c(\hx^k)+1]\}\nonumber\\
&\le&c(\hx^k)\log c(\hx^k)+\log
e+\frac{k(\log \beta)\log(2\beta)}{(1-\epsilon_k)\log
k}+\log[2\beta(k+1)]\nonumber\\
&\dfn&c(\hx^k)\log c(\hx^k)+k\cdot\varepsilon(k),
\end{eqnarray}
where we remind that $\beta$ is the cardinality of
$\hat{\calX}$, respectively, and where $\epsilon_k$ and
$\varepsilon(k)$ tend to zero as $k\to\infty$.

Our first lower bound on the compression ratio is conceptually very simple.
Since each $k$-block, $\hx_{ik+1}^{ik+k}$, $i=0,1,\ldots,n/k-1$, of the
reconstruction vector, $\hx^n$, is compressed using a finite-state machine
with $q$ states,
then according to \cite[Theorem 1]{ZL78}, its compression ratio is
lower bounded by
\begin{eqnarray} 
\frac{c(\hx_{ik+1}^{ik+k})+q^2}{k}\log\frac{c(\hx_{ik+1}^{ik+k})+q^2}{4q^2}&\ge&
\frac{c(\hx_{ik+1}^{ik+k})}{k}\log
c(\hx_{ik+1}^{ik+k})-\frac{c(\hx_{ik+1}^{ik+k})+q^2}{k}\log(4q^2)\nonumber\\
&\ge&\frac{c(\hx_{ik+1}^{ik+k})}{k}\log
c(\hx_{ik+1}^{ik+k})-\frac{(\log\beta)\log(4q^2)}{(1-\epsilon_k)\log
k}-\frac{q^2\log(4q^2)}{k},
\end{eqnarray} 
where the second inequality follows from \cite[eq.\ (6)]{ZL78}.
Since each $k$-block must comply with the
distortion constraint, this quantity is further lower bounded by
$$\min_{\hx^k\in\calB(x_{ik+1}^{ik+k},D)}\frac{c(\hx^k)}{k}\log
c(\hx^k)-\frac{(\log\beta)\log(4q^2)}{(1-\epsilon_k)\log
k}-\frac{q^2\log(4q^2)}{k},$$
and so, for the entire source vector $x^n$, we have
\begin{equation}
\rho(x^n;\calE(q,k,D))\ge\frac{1}{n}\sum_{i=0}^{n/k-1}
\min_{\hx^k\in\calB(x_{ik+1}^{ik+k},D)}c(\hx^k)\log c(\hx^k)-
\frac{(\log\beta)\log(4q^2)}{(1-\epsilon_k)\log
k}-\frac{q^2\log(4q^2)}{k}.
\end{equation}
For large enough $k$, the last two terms can be made arbitrarily small,
provided that $\log q\ll\log k$. Clearly, this lower bound can be
asymptotically attained by seeking the vector $\hx^k\in\hat{\calX^k}$ that
minimizes $c(\hx^k)\log c(\hx^k)$ across $\calB(x_{ik+1}^{ik+k},D)$ within
each $k$-block and compressing it by the LZ78 compression algorithm.

In order to state our second lower bound, we 
next define the joint empirical distribution of $\ell$-blocks of
$\hx_{ik+1}^{ik+k}$. Specifically, 
let $\ell$ divide $k$ which in turn divides $n$, and consider the
empirical distribution,
$\hat{P}_i=\{\hat{P}_i(\hx^\ell),~\hx^\ell\in\hat{\calX}^\ell\}$,
of $\ell$-vectors along the $i$-th $k$-block of $\hx^n$, which is
$\hx_{ik+1}^{ik+k}$,
$i=0,1,\ldots,n/k-1$, that is,
\begin{equation}
\hat{P}_i(\hx^\ell)=\frac{\ell}{k}\sum_{j=0}^{k/\ell-1}
\calI\{\hx_{ik+j\ell+1}^{ik+j\ell+\ell}=\hx^\ell\},~~~~\hx^\ell\in\hat{\calX}^\ell.
\end{equation}
Let $\hat{H}(\hX_i^\ell)$ denote the
empirical entropy of an auxiliary random $\ell$-vector, $\hX_i^\ell$, 
induced by $\hat{P}_i$, that is,
\begin{equation}
\hat{H}(\hX_i^\ell)=-\sum_{\hx^\ell\in\hat{\calX}^\ell}\hat{P}_i(\hx^\ell)\log
\hat{P}_i(\hx^\ell).
\end{equation}
Now, our second lower bound is given by the following inequality:
\begin{equation}
\label{2nd}
\rho(x^n;\calE(q,k,D))
\ge\frac{k}{n}\sum_{i=0}^{n/k-1}\min_{\hx_{ik+1}^{ik+k}\in\calB(x_{ik+1}^{ik+k},D)}
\frac{\hat{H}(\hat{X}_i^\ell)}{\ell}-
\frac{1}{\ell}\log\left\{q^2\left(1+\log\left[1+\frac{\beta^\ell}{q^2}
\right]\right)\right\}.
\end{equation}

\noindent
{\bf Discussion.}\\

Note that both lower bounds depend on the number of states, $q$, of the
FSLE, but not on the number of states, $|\calS|$, of the FSRE. In this sense,
no matter how large the number of states of the FSRE may be, none of these
bounds is affected. For the purpose of lower bounds, that establish
fundamental limitations, we wish to consider a class of encoders that is as broad as
possible, for the sake of generality. We therefore assume that $\calS$ is
arbitrarily large.\\

The second term on the right-hand side of 
(\ref{2nd}) is small when $\log q$ is small relative to $\ell$, which is in
turn smaller than $k$. This requirement is less restrictive than the parallel
one in the first bound, which was $\log q\ll \log k$. The bound is
asymptotically achievable by universal lossless coding of the vector
$\hx_{ik+1}^{ik+k}$ that minimizes $\hat{H}(\hX_i^\ell)$ within
$\calB(x_{ik+1}^{ik+k},D)$ using a universal lossless code that is based on
two-part coding: the first part is a header that indicates the type class
$\hat{P}_i$ using a logarithmic number of bits as a function of $k$ and the
second part is the index of the vector within the type class.\\

The main term of the second bound is essentially tighter than the main term of
the first bound since $\hat{H}(\hX_i^\ell)$ can be lower bounded by 
$c(\hx_{ik+1}^{ik+k})\log c(\hx_{ik+1}^{ik+k})$ minus some small terms (see,
e.g., \cite[eq.\ (26)]{me23}). On the other hand, the second bound is somewhat
more complicated due to the introduction of the additional parameter $\ell$.
It is not clear whether any one of the bounds completely
dominates the other one for any $x^n$. Anyway, it is always possible to choose the
larger bound between the two.\\

To prove eq.\ (\ref{2nd}), consider the following.
According to \cite[Lemma 2]{ZL78}, since the FSLE is an information lossless
encoder with 
$q$ states, it must obey the following generalized Kraft
inequality:
\begin{equation}
\sum_{\hat{x}^\ell\in\hat{\calX}^\ell}
2^{-\min_{z\in\calZ}L[f(z,\hat{x}^\ell)]}\le
q^2\left(1+\log\left[1+\frac{\beta^\ell}{q^2}\right]\right).
\end{equation}
This implies
that the description length at the output of the encoder is
lower bounded as follows.
\begin{eqnarray}
L(b^n)&=&\sum_{t=1}^n L[f(z_t,\hat{x}_t)]\nonumber\\
&=&\sum_{i=0}^{n/k-1}\sum_{m=0}^{k/\ell-1} \sum_{j=1}^\ell
L[f(z_{ik+m\ell+j},\hat{x}_{ik+m\ell+j})]\nonumber\\
&=&\sum_{i=0}^{n/k-1}\sum_{m=0}^{k/\ell-1}
L[f(z_{ik+m\ell+1},\hat{x}_{ik+m\ell+1}^{ik+m\ell+\ell})]\nonumber\\
&\ge&\sum_{i=0}^{n/k-1}\sum_{m=0}^{k/\ell-1}
\min_{z\in\calZ}L[f(z,\hat{x}_{ik+m\ell+1}^{ik+m\ell+\ell})]\nonumber\\
&=&\sum_{i=0}^{n/k-1}\frac{k}{\ell}\sum_{\hat{x}^\ell\in\hat{\calX}^\ell}\hat{P}_i(\hat{x}^\ell)
\cdot\min_{z\in\calZ}L[f(z,\hat{x}^\ell)],
\end{eqnarray}
Clearly,
\begin{equation}
\frac{L(b^n)}{n}\ge\frac{k}{n}\sum_{i=0}^{n/k-1}
\frac{1}{\ell}\sum_{\hat{x}^\ell\in\hat{\calX}^\ell}\hat{P}_i(\hat{x}^\ell)
\cdot\min_{z\in\calZ}L[f(z,\hat{x}^\ell)].
\end{equation}
Now, by the generalized Kraft inequality above,
\begin{eqnarray}
q^2\left(1+\log\left[1+\frac{\beta^\ell}{q^2}\right]\right)&\ge&
\sum_{\hat{x}^\ell\in\hat{\calX}^\ell}
2^{-\min_{z\in\calZ}L[f(z,\hat{x}^\ell)]}\nonumber\\
&\ge&\sum_{\hat{x}^\ell\in\hat{\calX}^\ell}
\hat{P}_i(\hat{x}^\ell)\cdot
2^{-\min_{z\in\calZ}L[f(z,\hat{x}^\ell)-\log
\hat{P}_i(\hat{x}^\ell)}\nonumber\\
&\ge&\exp_2\left\{-\sum_{\hat{x}^\ell\in\hat{\calX}^\ell}\hat{P}_i(\hat{x}^\ell)\cdot
\min_{z\in\calZ}L[f(z,\hat{x}^\ell)+\hat{H}(\hat{X}_i^\ell)\right\},
\end{eqnarray}
where the last inequality follows from the convexity of the exponential
function and Jensen's inequality.
This yields
\begin{equation}
\log\left\{q^2\left(1+\log\left[1+\frac{\beta^\ell}{q^2}\right]\right)\right\}\ge
\hat{H}(\hat{X}_i^\ell)-\sum_{\hat{x}^\ell\in\hat{\calX}^\ell}\hat{P}_i(\hat{x}^\ell)\cdot
\min_{z\in\calZ}L[f(z,\hat{x}^\ell),
\end{equation}
implying that
\begin{eqnarray}
\frac{L(b^n)}{n}&\ge&\frac{k}{n}\sum_{i=0}^{n/k-1}\frac{1}{\ell}\sum_{\hat{x}^\ell\in\hat{\calX}^\ell}
\hat{P}_i(\hat{x}^\ell)\cdot\min_{z\in\calZ}L[f(z,\hat{x}^\ell)]\nonumber\\
&\ge&\frac{k}{n}\sum_{i=0}^{n/k-1}\frac{\hat{H}(\hat{X}_i^\ell)}{\ell}-
\frac{1}{\ell}\log\left\{q^2\left(1+\log\left[1+\frac{\beta^\ell}{q^2}
\right]\right)\right\},
\end{eqnarray}
and since each $\hx_{ik+1}^{ik+k}$ must be in $\calB(x_{ik+1}^{ik+k}),D)$, the
summand of the first term on the left-hand side cannot be
smaller than
$\min_{\hx^k\in\calB(x_{ik+1}^{ik+k}),D)}\hat{H}(\hat{X}_i^\ell)/\ell$.
Since this lower bound on $L(b^n)/n$ holds for every $E\in\calE(q,k,D)$, it
holds also for $\rho(x^n;\calE(q,k,D))$.\\

Returning now to the first lower bound,
consider the following chain of inequalities:
\begin{eqnarray}
\label{chain}
\sum_{i=0}^{n/k-1}
\min_{\hx^k\in\calB(x_{ik+1}^{ik+k},D)}c(\hx^k)\log c(\hx^k)
&\ge&
\sum_{i=0}^{n/k-1}
\min_{\hx^k\in\calB(x_{ik+1}^{ik+k},D)}LZ(\hx^k)-k\varepsilon(k)\nonumber\\
&=&-\sum_{i=0}^{n/k-1}\log\left[\max_{\hx^k\in\calB(x_{ik+1}^{ik+k},D)}
2^{-LZ(\hx^k)}\right]-k\varepsilon(k)\nonumber\\
&\ge&-\sum_{i=0}^{n/k-1}\log\left[\sum_{\hx^k\in\calB(x_{ik+1}^{ik+k},D)}
2^{-LZ(\hx^k)}\right]-k\varepsilon(k).
\end{eqnarray}
It is conceivable that the last inequality may contribute most of the gap
between the left-most side and the right-most side of the chain (\ref{chain}),
since we pass from a single term in $\calB(x_{ik+1}^{ik+k},D)$ to the sum of all
terms in $\calB(x_{ik+1}^{ik+k},D)$. Since
\begin{equation}
\max_{\hx^k\in\calB(x_{ik+1}^{ik+k},D)}
2^{-LZ(\hx^k)}\le
\sum_{\hx^k\in\calB(x_{ik+1}^{ik+k},D)}
2^{-LZ(\hx^k)}\le
|\calB(x_{ik+1}^{ik+k},D)|\cdot\max_{\hx^k\in\calB(x_{ik+1}^{ik+k},D)}
2^{-LZ(\hx^k)},
\end{equation}
the gap between the left-most side
of (\ref{chain}) and the right-most side of (\ref{chain}) might take any positive value
that does not exceed $\log|\calB((x_{ik+1}^{ik+k},D)|$, which is in turn
approximately proportional to $k$ as $|\calB(x_{ik+1}^{ik+k},D)|$ is
asymptotically exponential in $k$. Thus, the right-most side of (\ref{chain}),
corresponds to a coding rate which might be strictly smaller than that of the
left-most side. 
Yet, we argue that the right-most side of
(\ref{chain}) can still be asymptotically attained by a finite-state encoder.
But to this end, its FSLE
component should possess $q=\beta^k$ states, as it is actually a block code of length $k$.
In order to see this, we need to define the following
{\em universal probability distribution} (see
also \cite{me23}, \cite{CM21}, \cite{MC20}):
\begin{equation}
\label{Udis}
U(\hx^k)=\frac{2^{-LZ(\hx^k)}}{\sum_{\tx^k\in\hat{\calX}^k}2^{-LZ(\tx^k)}}
\dfn \frac{2^{-LZ(\hx^k)}}{Z},~~~~\hx^k\in\hat{\calX}^k,
\end{equation}
and accordingly, define also
\begin{equation}
U[\calB(x^k,D)]=\sum_{\hx^k\in\calB(x^k,D)}U(\hx^k).
\end{equation}
Now, the first term on the right-most side of (\ref{chain}) can be further manipulated as
follows.
\begin{eqnarray}
\label{last}
-\sum_{i=0}^{n/k-1}\log\left[\sum_{\hx^k\in\calB(x_{ik+1}^{ik+k},D)}
2^{-LZ(\hx^k)}\right]&=&
-\sum_{i=0}^{n/k-1}\log\left[\sum_{\hx^k\in\calB(x_{ik+1}^{ik+k},D)}
U(\hx^k)\cdot Z\right]\nonumber\\
&\ge&-\sum_{i=0}^{n/k-1}\log U[\calB(x_{ik+1}^{ik+k},D)],
\end{eqnarray}
where the last inequality is because $-\log Z\ge 0$ thanks to Kraft's
inequality applied to the code-length function $LZ(\cdot)$.

Now, the last expression in (\ref{last})
suggests achievability using the universal distribution, $U$, for independent random
selection of the various codewords. The basic idea is quite standard and
simple: The quantity $U[\calB(x_{ik+1}^{ik+k},D)]$ is the
probability that a single randomly chosen reproduction vector, drawn
under $U$, would fall within
distance $kD$ from the source vector, $x_{ik+1}^{ik+k}$. If all reproduction
coedwords are drawn independently under $U$, then the typical number of such
random selections that it takes before one sees the first one in
$\calB(x_{ik+1}^{ik+k},D)$, is of the
exponential order of $1/U[\calB(x_{ik+1}^{ik+k},D)]$. Given that the codebook
is revealed
to both the encoder and decoder, once it has been selected, the encoder merely
needs to transmit
the index of the first such reproduction vector within the
codebook, and the description length of that index can be made essentially as
small as
$\log\{1/U[\calB(x_{ik+1}^{ik+k},D)]\}=-\log(U[\calB(x_{ik+1}^{ik+k},D)])$.
In \cite{me23}, we use this simple idea to prove achievability for an
arbitrary distortion measure. More precisely, the following theorem is stated
and proved
in \cite{me23} with some adjustment of the notation:\\

\noindent
{\bf Theorem (\cite[Theorem 2]{me23}).}
Let $d:\calX^k\times\hat{\calX}^k\to\reals^+$ be an arbitrary distortion
function. Then, for every $\epsilon>0$, there
exists a sequence of $d$-semifaithful, variable-length block codes of block
length $k$,
such that for every
$x^k\in\calX^k$, the code length for $x^k$ is upper bounded by
\begin{equation}
L(x_{ik+1}^{ik+k})\le-\log(U[\calB(x_{ik+1}^{ik+k},D)])+(2+\epsilon)\log k +c+\delta_k,
\end{equation}
where $c> 0$ is a constant and $\delta_k=O(kJ^ne^{-k^{1+\epsilon}})$.\\

Through the repeated application of this code for each one of the $n/k$ blocks
of length of $k$, the
lower bound of the last line of (\ref{last}) is asymptotically attained. As elaborated
on in \cite{me23}, the ensemble of codebooks selected under the universal
distribution $U$ exhibits universality in both the source sequence slated for
encoding and the chosen distortion measure. This stands in contrast to the
classical random coding distribution, which typically relies on both the
statistics of the source and the characteristics of the distortion measure.\\

\noindent
{\bf Discussion.}\\

A natural question that may arise is whether this performance is the best that
can be attained given that the number of FSLE states, $q$, is as large as
$\beta^k$. For now, this question remains open, but it is conjectured that the
answer is affirmative, in view of the matching converse theorem of
\cite[Theorem 1]{me23}, which applies to the vast majority of source sequences in each
and every type class of any order, and even without the limitation 
of finite-state encoders.\\

It is natural to think of the memory resource used by a finite-state
encoder in terms of the number of bits (or equivalently, 
the size of of a register) needed in order to store the current state at each time
instant, namely, the base 2 logarithm of the total number of states. Indeed, both
lower bounds derived earlier contain terms that are proportional to $\log q$,
the memory size pertaining to the FSLE. Since the memory size, $\log|\calS|$
of the FSRE is assumed arbitrarily large, as discussed earlier, the total
size of the encoder memory, $\log|\calS|+\log q$, is dominated by
$\log|\calS|$, and so, the contribution of $\log q$ to the total memory volume can be considered
negligibly small. Therefore, one of our main 
messages in this work is that, as far as the total memory size goes, 
it makes very little difference if we allow $\log
q$ to be as large as $k\log\beta$ and thereby achieve the better performance,
rather than keeping $\log q$ smaller and then ending up with the inferior
compression performance of minimizing $LZ(\hx_{ik+1}^{ik+k})$ within
$\calB(x_{ik+1}^{ik+k},D)$ for each block.\\

\section{Conclusion}

In this paper, we revisited the paradigm of lossy compression of individual
sequences using finite-state machines, as a natural extension to the same paradigm
in the lossless case, as established by Ziv and Lempel in \cite{ZL78} and
other related works. This work can also be viewed as revisit of \cite{me23}
from the perspective of finite-state encoding of individual sequences.
Our model of a finite-state encoder is that of a cascade
of the finite-state $k$-dimensional reproduction encoder (with an arbitrarily
large number of states) and a finite-state lossless encoder,
acting on the reproduction sequence.
Our main contributions in this work are the following.
\begin{enumerate}
\item We derived two different lower bounds to the compression ratio.
\item We have shown that both bounds depend on the number of states, $q$, of the lossless
encoder, but not on the number of states of the reproduction encoder.
\item We have shown that for relatively small $q$, one cannot do better than
seeking the most compressible reproduction sequence within the 'sphere' of
radius $kD$ around the source vector. Nonetheless, if we allow $q=\beta^k$, we can improve
performance significantly by using a good code from the ensemble of codes
where each codeword is selected independently at random under the universal
distribution, $U$. The resulting code is universal, not only in the sense of
the source sequence, as in \cite{ZL78}, but also in the distortion function,
in the sense discussed in \cite{me23}.
This passage from small $q$ to large $q$ will not increase
the total memory resources of entire encoder significantly, considering the
large memory  that may be used by the reproduction encoder anyway.
\item We suggest the conjecture that the performance achieved as described in item 3 is
the best performance achievable for large $q$.
\end{enumerate}

Finally, we comment that our derivations can be extended to incorporate side
information, $u^n=(u_1,u_2,\ldots,u_n)$, available to
both the encoder and decoder. In the model of the finite-state encoder, this
amounts to allowing both the FSRE and the FSLE sequential access
to $u_t$, $t=1,2,\ldots$. The decoder, of course, should also have access to
$u^n$. Another modification needed is in replacing the LZ algorithm by its
conditional version in all places (see, e.g., \cite{me22},
\cite{me20}, and \cite{Ziv85}).

\end{document}